\documentclass[conference]{IEEEtran}
\IEEEoverridecommandlockouts
\usepackage{cite}
\usepackage{amsmath,amssymb,amsfonts}
\usepackage{algorithmic}
\usepackage{graphicx}
\usepackage{textcomp}
\usepackage{xcolor}
\usepackage{amsthm}
\usepackage{overpic}
\usepackage{bm}
\usepackage{overpic}
\usepackage{float}
\usepackage{amsthm}
\usepackage{algorithm,algorithmic}

\theoremstyle{plain}

\newtheorem{lemma}{Lemma}

\DeclareMathOperator{\sinc}{sinc}
\def\CN{\mathcal{N}_{\mathbb{C}}} 
\makeatletter
\newcommand\fs@ruled@notop{\def\@fs@cfont{\bfseries}\let\@fs@capt\floatc@ruled
  \def\@fs@pre{}%
  \def\@fs@post{\kern2pt\hrule\relax}%
  \def\@fs@mid{\kern2pt\hrule\kern2pt}%
  \let\@fs@iftopcapt\iftrue}
\renewcommand\fst@algorithm{\fs@ruled@notop}
\makeatother

\def\CN{\mathcal{N}_{\mathbb{C}}} 

\begin{document}

\title{Control Signaling for Reconfigurable Intelligent Surfaces: How Many Bits are Needed?
}
\author{\thanks{This work was supported by the FFL18-0277 grant from the Swedish Foundation for Strategic Research.}

\IEEEauthorblockN{Anders Enqvist$^*$, {\"O}zlem Tu\u{g}fe Demir$^\dagger$, Cicek Cavdar$^*$, and  Emil Bj{\"o}rnson$^*$}
\IEEEauthorblockA{ {$^*$Department of Computer Science, KTH Royal Institute of Technology, Kista, Sweden}
\\ {$^\dagger$Department of Electrical-Electronics Engineering, TOBB University of Economics and Technology, Ankara, Türkiye
		} \\
		{Email: enqv@kth.se, ozlemtugfedemir@etu.edu.tr,  cavdar@kth.se, emilbjo@kth.se }
}

}


\maketitle

\begin{abstract}
Reconfigurable intelligent surfaces (RISs) can greatly improve the signal quality of future communication systems by reflecting transmitted signals toward the receiver. However, even when the base station (BS) has perfect channel knowledge and can compute the optimal RIS phase-shift configuration, implementing this configuration requires feedback signaling over a control channel from the BS to the RIS. This feedback must be kept minimal, as it is transmitted wirelessly every time the channel changes.
In this paper, we examine how the feedback load, measured in bits, affects the performance of an RIS-aided system. Specifically, we investigate the trade-offs between codebook-based and element-wise feedback schemes, and how these influence the signal-to-noise ratio (SNR). We propose a novel quantization codebook tailored for line-of-sight (LoS) that guarantees a minimal SNR loss using a number of feedback bits that scale logarithmically with the number of RIS elements.
We demonstrate the codebook's usefulness over Rician fading channels and how to extend it to handle a non-zero static path. Numerical simulations and analytical analysis are performed to quantify the performance degradation that results from a reduced feedback load, shedding light on how efficiently RIS configurations can be fed back in practical systems.
\end{abstract}


\section{Introduction}
Reconfigurable intelligent surfaces (RISs) have emerged as a promising new hardware entity that can improve the signal coverage of future cellular networks without requiring more base stations (BSs). This feature is particularly important when networks are deployed at higher frequencies in future networks because the coverage otherwise shrinks. An RIS consists of many controllable reflecting elements. By dynamically adjusting the phase shifts of the electromagnetic waves impinging on these elements, the RIS can optimize the shape of the reflected signal. This feature is especially useful in propagation environments without a line-of-sight (LoS) path between the transmitter and receiver \cite{RIS-EE,bjornson2022reconfigurable}. The phase shifts at the RIS must be adapted to channel variations and updated rapidly without consuming too many radio resources.

Much of the existing research has focused on estimating the channel at the BS and user equipment (UE) and using those estimates at the BS to compute the preferred RIS configuration. These efforts aim to minimize the overhead associated with channel estimation to improve spectral or energy efficiency; see \cite{Renzo2020a} for a review of fundamental results. A significant gap in the current literature is the lack of methods to transfer the phase-shift configurations to the RIS efficiently. A control channel is needed to rapidly convey the configuration \cite{bjornson2022reconfigurable}.

The control channel between the BS and RIS should be wireless since the main point of using an RIS instead of an extra BS is that it does not need a wired backhaul link. Previous works assume the configuration can be transmitted from the BS to the RIS without requiring any resources. In practical systems, such as 5G, control channels either have limited dedicated resources or share resources with the data channel \cite{dahlman20235g}. Hence, to make RIS practical, it is essential to develop resource-efficient configuration feedback mechanisms.

In this paper, we analyze how the phase-shift configuration feedback from the BS to the RIS impacts the overall system performance. Specifically, we develop a novel codebook and examine how its size---measured in bits per reconfiguration---affects the signal-to-noise ratio (SNR) metric. 

\subsection{Related work}

In \cite{stamatelis2024multi}, a neural network was designed to jointly optimize the BS and RIS precoders, aiming to reduce information exchange overhead. The most closely related paper is \cite{saggese2024impact}, which proposes and evaluates the impact of RIS control operations based on beam sweeping or channel estimation, assessing their suitability for mobility scenarios and control data exchange. In contrast, in this paper, we provide a detailed analysis of the actual number of feedback bits required to achieve a particular SNR and also consider the effects of multiple BS antennas and the impact of different channel models (including a static path between the UE and BS) on the feedback design. The overhead associated with the acquisition of channel state information (CSI) and the subsequent feedback of the optimized configuration to the RIS is considered in \cite{zappone2020overhead}.
Similarly, \cite{saggese2023control} analyzed the impact of the RIS control channel on resource allocation in mobile edge computing. We also acknowledge the valuable insights collected within the scope of the Bachelor thesis \cite{miranda2024reconfigurable} supervised by the first author.

\subsection{Motivation and contribution}

In contrast to prior work, we provide concrete figures that quantify the number of bits required in an RIS configuration feedback channel to achieve a minimal SNR loss in a LoS scenario. To maintain performance when beamforming towards a specific UE, channel estimation and RIS configurations must be updated once per coherence time \cite{jian2022reconfigurable}; thus, few resources can be dedicated to feedback. For pure LoS scenarios, we propose a codebook where the desired configuration can be indicated using a number of bits proportional to the logarithm of the number of RIS elements. The logarithmic scaling is faster than the linear scaling obtained with a naive element-wise quantization and is crucial given the large number of elements in typical RISs. 
Note that ensuring error-free control channels may require additional channel coding, which significantly increases the overhead, as seen in 5G systems \cite{3gpp.38.212}.

This paper aims to address the following key questions:

\begin{itemize} 
\item How does the RIS configuration feedback size affect the SNR compared to an ideal configuration?
\item What are the performance trade-offs between codebook-based and element-wise feedback schemes, and how does the channel model affect it?

\end{itemize}

Through numerical simulations, we quantify the performance degradation caused by reducing the feedback size, offering insight into how the control signaling overhead can be efficiently managed in RIS-aided communication systems.

\begin{figure}[t!]
	\centering 
	\begin{overpic}[width=.99\columnwidth,tics=10]{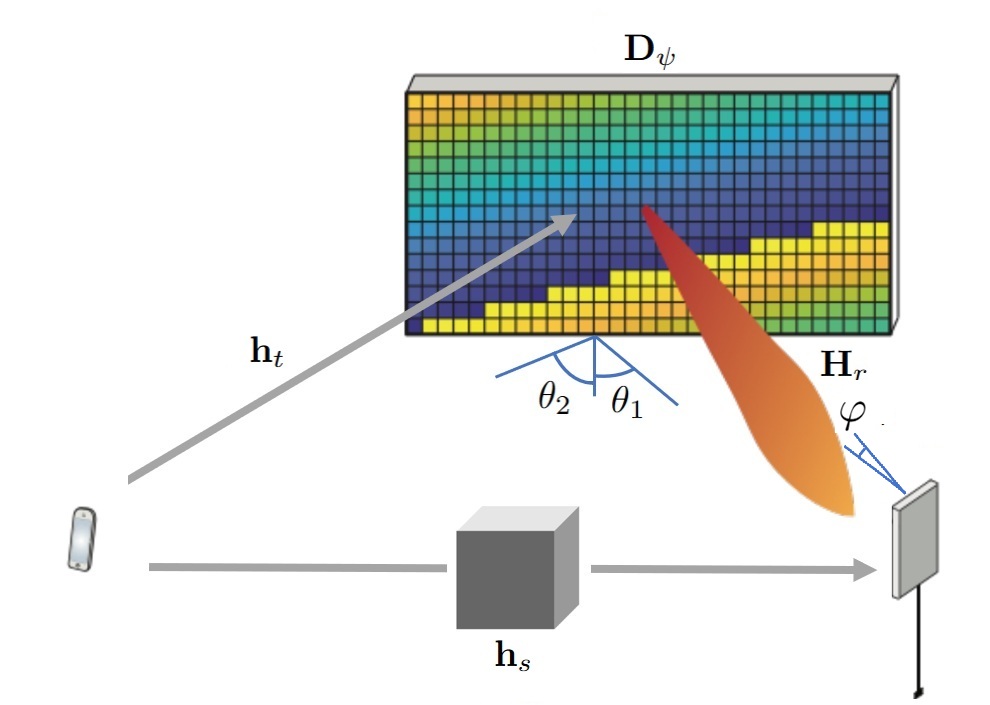}
		\put(94,9.5){\footnotesize BS}
		\put(2,9.5){\footnotesize  UE}
\end{overpic}  \vspace{-2mm}
	\caption{We consider a setup where a UE transmits to a BS through a static path $\mathbf{h}_s$ and a controllable RIS with $N$ elements. The channel from the BS to the RIS is in LoS with an angle $\theta_1$ at the RIS and the channel from the RIS to the UE is in LoS with an angle $\theta_2$. The BS receives the signal from the RIS at an angle $\varphi$. The colors represent the phase shifts of the RIS elements. }  \vspace{-3mm}
\label{fig:system-model}
\end{figure}

\section{System Model} \label{sec:systemmodel}

We consider the uplink system illustrated in Fig.~\ref{fig:system-model}, where a single-antenna UE transmits to a multi-antenna BS equipped as a uniform linear array (ULA) with $K$ antennas. The communication is aided by an RIS that is also modeled as a ULA with $N$ elements.\footnote{The results are also applicable to uniform planar arrays when the BS, UE, and RIS are located in the same plane. The three-dimensional extension will be considered in future work.}  All elements in the RIS and the BS are assumed to be at an inter-element spacing of half a wavelength.

The end-to-end channel vector $\mathbf{h} \in \mathbb{C}^{K}$ is expressed as
\begin{equation} \label{eq:end-to-end}
    \mathbf{h}=\mathbf{h}_s + \mathbf{H}_r \mathbf{D}_\psi \mathbf{h}_t,
\end{equation}
where $\mathbf{h}_s \in \mathbb{C}^{K}$ is the static path between the BS and the UE, $\mathbf{H}_r \in \mathbb{C}^{K \times N}$ is the channel from the
BS to the RIS, and $\mathbf{h}_t \in \mathbb{C}^{N}$ is the channel from the RIS to
the UE. The diagonal phase-shift matrix $\mathbf{D}_\psi \in \mathbb{C}^{N \times N}$ that specifies the RIS configuration is defined as 
\begin{equation}
    \mathbf{D}_\psi=\mathrm{diag}(e^{j \psi_1 } , \dots , e^{j \psi_N }),
\end{equation}
where $\psi_n \in [-\pi,\pi)$ denotes the phase shift induced by element $n$ before the incident signal is reflected.  

The SNR for data transmission with a given phase-shift configuration, when transmitting with power $P$ and having the receiver noise power $\sigma^2$, is given by 
\begin{equation}
    \mathrm{SNR}=\frac{P}{\sigma^2}\Vert\mathbf{h}\Vert^2.
\end{equation}
For given realizations of $\mathbf{h}_s$, $\mathbf{H}_r$, and $\mathbf{h}_t$, the BS can compute the RIS configuration that maximizes SNR, for instance, by following the algorithm in \cite{zhang2020capacity}. We do not consider the channel estimation procedures in this paper because previous work has covered it in detail.
Instead, we assume that the BS has determined a suitable RIS configuration and needs to convey it to the RIS through the transmission of a feedback message consisting of $t$ bits. This message is sent over a control channel with orthogonal time-frequency resources and is assumed to be coded such that the reception is free of errors.

\subsection{Naive element-wise quantization}

The naive approach to configuration feedback is to quantize each phase value in $\mathbf{D}_\psi$ separately using $b$ bits so that the configuration is represented by $t=Nb$ bits in total. This leads to a feedback load that grows linearly with the number of RIS elements, which is problematic since practical implementations of RIS can feature hundreds or even thousands of elements \cite{pei2021ris}. In the following section, we develop a codebook that requires less feedback by exploiting typical channel characteristics. We also examine the performance of element-wise quantization analytically.

\section{Codebook Design for LoS Channels}

It is known that RISs are particularly useful for creating virtual LoS paths between transmitters and receivers \cite{bjornson2022reconfigurable}. In this section, we consider the codebook design tailored for such scenarios.
We assume the static path is negligible to focus the analysis on the cascaded channel through the RIS, and will then extend the codebook design in Section~\ref{sec:with-static-path}. We assume the RIS has LoS to both the BS and the UE. As illustrated in Fig.~\ref{fig:system-model}, the angle of arrival of the incident wave from the UE to the RIS is $\theta_2$, while the angle of departure from the RIS towards the BS is $\theta_1$. The corresponding angle of arrival at the BS is $\varphi$. This allows us to model $\mathbf{H}_r$ and $\mathbf{h}_t$ using array response vectors. We begin by defining these array response vectors for $i=1,2$ as
\begin{equation}
    \mathbf{a}_N(\theta_i)=\left[1, \; e^{-j \pi \sin \theta_i}, \dots, e^{-j \pi (N-1)\sin \theta_i}\right]^T,
\end{equation}
\begin{equation}
    \mathbf{a}_K(\varphi)=\left[1, \; e^{-j \pi \sin \varphi}, \dots, e^{-j \pi (K-1)\sin \varphi}\right]^T.
    \vspace{-1mm}
\end{equation}
Then, we can write the end-to-end channel in \eqref{eq:end-to-end} as
    \vspace{-0.6mm}
\begin{equation}
    \mathbf{h}=\sqrt{\beta_r \beta_t}\mathbf{a}_K(\varphi)  \mathbf{a}_N^T(\theta_1) \mathbf{D}_\psi \mathbf{a}_N(\theta_2),     \vspace{-0.6mm} 
\end{equation}
where  $\mathbf{h}_t=\sqrt{\beta_t}\mathbf{a}_N(\theta_2)$ and $\mathbf{H}_r= \sqrt{\beta_r}\mathbf{a}_K(\varphi)\mathbf{a}_N^T(\theta_1)$. We assume that the RIS is located in the far-field region of the BS and the UE, so the average channel gains $\beta_r$ and $\beta_t$ from each RIS element to the BS or the UE are the same. 

The SNR is proportional to the array gain $\mathrm{G_A}$, which is the squared magnitude of the array factor of the RIS given by 
\vspace{-0.6mm}
\begin{align}{\mathrm{G_A}}&=\left\vert\mathbf{a}_N^T(\theta_1) \mathbf{D}_\psi \mathbf{a}_N(\theta_2) \right\vert^2 \nonumber \\
    &= \left\vert\sum_{n=1}^{N} e^{-j (\pi(n-1)(\sin\theta_1+\sin\theta_2)-\psi_n)}\right\vert^2.
\end{align}
In this setup, it is maximized if we select the RIS phase shifts 
\begin{equation} \label{eq:optimal-phases}
    \psi_n = \pi (n-1) (\sin{\theta_1} + \sin{\theta_2})
\end{equation}
for $n=1,\ldots,N$, which is equivalent to setting  
\begin{equation}
    \mathbf{D}_\psi=\mathrm{diag} \left( \mathbf{a}_N^*(\theta_1) \odot \mathbf{a}_N^*(\theta_2) \right)
    \label{eq:D_psi},
\end{equation}
where $\odot$ denotes the element-wise product. This configuration gives the maximum array gain ${\mathrm{G_A}}=N^2$. Hence, for this type of channel, we achieve the maximal channel gain
\begin{equation}
     \Vert\mathbf{h}\Vert^2= N^2 K \beta_r \beta_t
\end{equation}
when using maximum-ratio combining (MRC) at the BS.

We assume that the BS and the RIS controller have a common codebook, $C = \{\mathbf{D}_1,\ldots,\mathbf{D}_{|C|}\}$, with RIS configurations. The BS can then convey the desired configuration by sending the index of a codebook entry using $\lceil \log_2(|C|) \rceil$ bits of feedback information. For a given number of feedback bits, $l$,
the problem is to design a codebook with $|C|=2^l$ entries that gives the highest communication performance. We will first identify the codebook structure and then analyze how large the cardinality $|C|$ must be to have a minimal SNR loss.

\subsection{Codebook design for pure LoS channels}

The optimal phase shifts in \eqref{eq:optimal-phases} depend on the element index $n$ and on the value $(\sin{\theta_1} + \sin{\theta_2})$.
Hence, it is sufficient for the BS to convey the latter value to the RIS, which can then compute all the $N$ phase shifts.
However, since the control channel has limited resources, the BS must quantize the value to finite precision while guaranteeing that the performance loss is acceptable. Suppose some nearby value $\hat{\Theta}\approx \sin{\theta_1}+\sin{\theta_2}$ is conveyed to the RIS, which then uses the configuration $\psi_n = \pi (n-1) \hat{\Theta}$. In this case, the array gain becomes 
\begin{align} \nonumber
  {\mathrm{G_A}}&= \left\vert\sum_{n=1}^{N} e^{-j \pi\left(n-1\right)\left(\sin\theta_1+\sin\theta_2-\hat{\Theta}\right)}\right\vert^2 \\
    &=\frac{\left|1-e^{-j N\pi\left(\sin\theta_1+\sin\theta_2-\hat{\Theta}\right)}\right|^2}{\left|1-e^{-j \pi\left(\sin\theta_1+\sin\theta_2-\hat{\Theta}\right)}\right|^2} \\ 
    &\approx N^2 \sinc^2 \left(  \frac{N\left(\sin\theta_1+\sin\theta_2-\hat{\Theta}\right)}{2}\right).
\end{align}
To derive this, we used the summation formula for geometric series, Euler's formula, and that $\sin^2(x)\approx x^2$ for small values of $x$. This is similar to beamforming analysis in \cite[Sec.~4.3]{bjornson2024introduction}. As the difference between $\hat{\Theta}$ and $\sin\theta_1+\sin\theta_2$ increases, the array gain becomes smaller than the ideal value $N^2$.

In practice, the potential UEs and BSs can be located at any angle, allowing both $\theta_1$ and $\theta_2$ to vary between $(-\pi/2,\pi/2)$. It then follows that $\sin \theta_1 + \sin \theta_2$ attains values in the interval $(-2,2)$. Using $l$ bits in the control channel for feedback, we can divide this interval into $|C|=2^l$ parts of equal length, with each part represented by its midpoint value $\Theta_i$ for $i=1,\dots,|C|$.

The deviation of $\hat{\Theta}$ from its optimum results in a lower array gain. 
When analyzing array gains, the half-power beamwidth (HPBW) is commonly used as the interval within which losses are considered acceptable. It is defined as the range over which the array gain, ${\mathrm{G_A}}$, is at least half of its maximum value, i.e., $N^2/2$. The HPBW is approximately $\mathrm{HPBW}(N)=\frac{1.772}{N}$ 
\cite[Sec. 4.3]{bjornson2024introduction}. The HPBW requirement can be formulated as

\begin{equation}
    2\left|\sin{\theta_1}+\sin{\theta_2}-\hat{\Theta}\right| \leq \mathrm{HPBW}(N),
\end{equation}
which implies that we must ensure 
\begin{equation}
    \frac{|(-2,2)|}{2^l} \leq \frac{1.772}{N} \iff l \geq \log_2\left(\frac{4N}{1.772}\right),
\end{equation}
which is guaranteed for
\begin{equation}
    l=\lceil 1.1746+\log_2(N) \rceil 
    \label{eq:b_HPBW}.
\end{equation}
This design ensures that the array gain provided by the RIS remains above $N^2/2$, regardless of what angles $\theta_1,\theta_2$ we happen to come across in practice. 
With this design, the entry $\mathbf{D}_i$ in the codebook is given by
\begin{equation} \label{eq:precoder}
   \mathbf{D}_i = \mathrm{diag} \left( 1, e^{j\pi \Theta_i},\ldots,e^{j\pi(N-1)\Theta_i}  \right),
\end{equation} 
where $\Theta_i=-2+2^{1-l}+(i-1)2^{2-l}$ for $i=1,\ldots,2^l$.

It is noteworthy that \eqref{eq:b_HPBW} has a logarithmic scaling with the number of RIS elements. This can be compared to the linear scaling obtained in the naive element-wise quantization described in Section \ref{sec:systemmodel}-A. The improved scaling is obtained for exploiting the structure of LoS channels.

\subsection{Extension to Rician fading channels}

In practical LoS channels, diffuse scattering exists in addition to the direct LoS path. This is particularly the case for the channel between the UE and RIS, because the UE will typically have scattering objects in its vicinity. Hence, we consider the case when the vector $\mathbf{h}_t$ experiences Rician fading with the factor $\kappa$. This implies that the channel is defined as
\begin{equation}
    \mathbf{h}_t=\sqrt{\frac{\kappa \beta_t}{\kappa+1}}\mathbf{a}_N(\theta_2)+\sqrt{\frac{\beta_t}{\kappa+1}}\mathbf{w}_N,
\end{equation}
\begin{figure*}
\begin{align}
    \Vert\mathbf{h}\Vert^2&= \left \Vert \sqrt{\beta_r \beta_t}\mathbf{a}_K(\varphi)  \mathbf{a}_N^T(\theta_1) \mathbf{D}_\psi   \left( \sqrt{\frac{\kappa}{\kappa+1}}\mathbf{a}_N(\theta_2)+\sqrt{\frac{1}{\kappa+1}}\mathbf{w}_N\right) \right \Vert^2 \nonumber\\
    &= \beta_r\beta_t\left \Vert \mathbf{a}_K(\varphi)   \sqrt{\frac{\kappa}{\kappa+1}}N+\sqrt{\frac{1}{\kappa+1}}\mathbf{a}_K(\varphi)\mathbf{a}_N^T(\theta_1) \mathbf{D}_\psi\mathbf{w}_N \right  \Vert^2 =K\beta_r\beta_t\left|    \sqrt{\frac{\kappa}{\kappa+1}}N+\sqrt{\frac{1}{\kappa+1}} \sum_{n=1}^{N}\tilde{w}_n \right|^2. \label{eq:channel-gain} 
\end{align}
\hrulefill \vspace{-4mm}
\end{figure*}
where $\mathbf{w}_N\sim \CN(\mathbf{0},\mathbf{I}_N)$  and the factor $\kappa$ determines how much power is delivered by the LoS and the non-LoS path, respectively, with $\kappa \to \infty$ representing the pure LoS case considered earlier and $\kappa \to 0$ implying pure Rayleigh fading. The 3GPP model in \cite{3gpp.25.996} specifies the Rician factor as $\kappa=13-0.03m$ [dB], where $m$ is the distance between the UE and RIS in meters. In the simulations, we let $\kappa=10$\,dB corresponding to a UE located 100\,m  from the RIS.

Suppose the phase shifts are selected as in  \eqref{eq:D_psi} by only considering the LoS path. The channel gain becomes as in \eqref{eq:channel-gain} as shown at the top of the page. In that expression, the coefficients $\tilde{w}_n$ have the same distribution as the entries of $\mathbf{w}_N$. The average channel gain then becomes
\vspace{-0.6mm}
\begin{equation} \label{eq:average-Rician}
    \mathbb{E}\left\{ \Vert\mathbf{h}\Vert^2 \right\}=
    \frac{\kappa}{\kappa+1}K N^2 \beta_r \beta_t + \frac{1}{\kappa+1}K N \beta_r \beta_t. \vspace{-0.6mm}
\end{equation}
This result has a logical physical explanation: we maintain the full array gain of $N^2$ for the LoS component. However, for the non-LoS component, the array gain is reduced to $N$ since the RIS can only capture additional energy proportional to its size ($N$) but not beamform it towards the BS. When $\kappa$ is large, the first term in \eqref{eq:average-Rician} dominates; thus, we can use the proposed codebook with only a minor performance loss.

\vspace{-0.6mm}
\subsection{Comparison with element-wise quantization}
\vspace{-0.6mm}
The alternative feedback approach is to quantize each element's phase shift individually without relying on specific channel conditions. The quantization grid for each element will then consist of equally-spaced points on the unit circle. The following lemma quantifies the array gain in this case.

\begin{lemma} \label{lemma1}
Suppose the distribution of the phase-shift $\psi_n$ for RIS element $n$ is uniform on $[-\pi,\pi)$.
If $2^b$ equally-spaced quantization points on the unit circle are used and the quantization errors $\tilde{\psi}_n$ and $\tilde{\psi}_m$ are independent for $n\neq m$, then when using $b$ quantization bits per element, the array gain is degraded to %
\vspace{-0.6mm}
\begin{equation} \label{eq:EGA}
    \mathbb{E}\{\mathrm{G_A}\} = N^2\sinc^2\left(\frac{1}{2^b}\right)+N\left(1-\sinc^2\left(\frac{1}{2^b}\right)\right). \vspace{-0.6mm} 
\end{equation}  
\end{lemma}

\begin{IEEEproof}
    
    With uniformly quantization of the optimal phase shifts in \eqref{eq:D_psi}, we obtain the configuration $Q_b(\mathbf{D}_\psi)=Q_b(\mathrm{diag} \left( \mathbf{a}_N^*(\theta_1) \odot \mathbf{a}_N^*(\theta_2)\right))$ where $Q_b(\cdot)$ is the $b$-bit quantization operator. We then obtain the average array gain
    \vspace{-0.6mm}
    \begin{align}
        \mathbb{E}\{\mathrm{G_A}\}&=\mathbb{E}\left\{\left|\mathbf{a}_N^T(\theta_1)Q_b(\mathbf{D}_\psi)\mathbf{a}_N(\theta_2)\right|^2\right\}. 
    \end{align}

Since $2^b$ equally-spaced quantization points on the unit circle are used, the quantization error $\tilde{\psi}_n$ is uniformly distributed in $\left(-\pi/2^b,\pi/2^b\right)$.  Hence, the moment generating function of $\tilde{\psi}_n$ is given as
\vspace{-0.6mm}
\begin{align}
    \mathbb{E}\left\{ e^{j \tilde{\psi}}\right \}&=\int_{-\pi/2^b}^{\pi/2^b} \frac{1}{2\pi/2^b}e^{j u}du= \frac{e^{j\pi/2^b}-e^{-j \pi/2^b}}{2j(\pi/2^b)}\nonumber\\
    &=\sinc \left(\frac{1}{2^b}\right), \label{eq:MGF}
\end{align}
we obtain
\vspace{-0.6mm}
\begin{align}
        \mathbb{E}\{\mathrm{G_A}\}&=\mathbb{E}\left\{\left|\sum_{n=1}^N e^{j \tilde{\psi}_n} \right|^2\right\}  =\mathbb{E}\left\{\sum_{n=1}^N \sum_{m=1}^N e^{j (\tilde{\psi}_n-\tilde{\psi}_m)} \right\} \nonumber \\
        &=N+N(N-1)\mathbb{E}\left\{  e^{j \tilde{\psi}_n}   \right\}\mathbb{E}\left\{  e^{-j \tilde{\psi}_m}   \right\} \nonumber \\
        &=N^2\sinc^2\left(\frac{1}{2^b}\right)+N\left(1-\sinc^2\left(\frac{1}{2^b}\right)\right) . 
    \end{align}    
\end{IEEEproof}

\begin{figure}[t!] \centering \begin{overpic}[width=.99\columnwidth,tics=10]{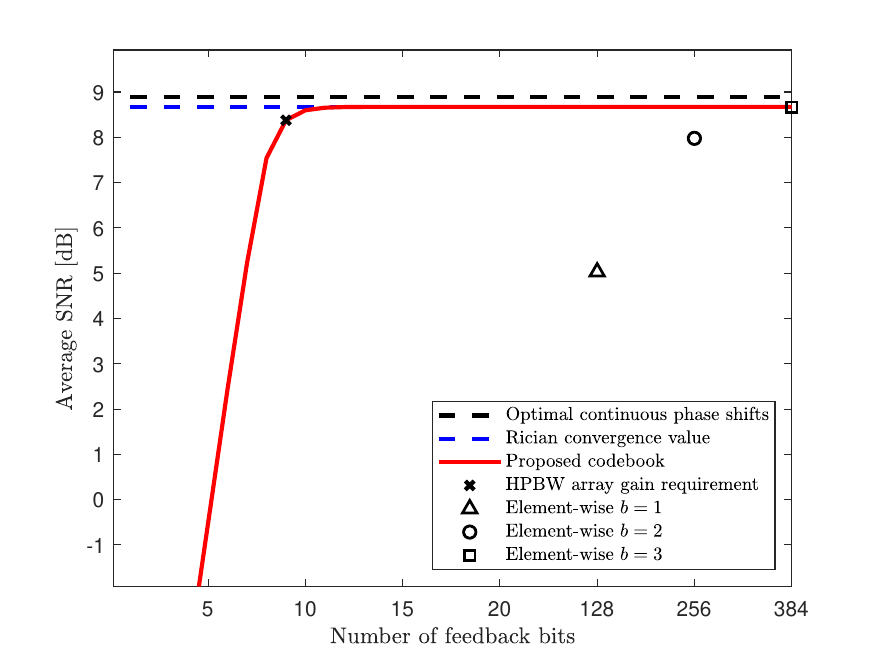}
		\put(79,60.5){\footnotesize $t=384$}
		\put(67,58.5){\footnotesize $t=256$}
            \put(55,43.5){\footnotesize $t=128$}
\end{overpic}  \vspace{-8mm}
\caption{The uplink SNR that is achieved with an RIS with 128 elements under Rician fading with $\kappa=10$\,dB, when the proposed codebook or element-wise quantization is used.}
\label{fig:Rice_fading}
\vspace{-3mm}
\end{figure}

An approximation
\vspace{-0.6mm}
\begin{equation}
    \mathbb{E}\{\mathrm{G_A}\} \approx N^2\sinc^2\left(\frac{1}{2^b}\right), \vspace{-0.6mm}
\end{equation}
can also be obtained by dropping the last term in \eqref{eq:EGA}, which becomes tighter as $N$ increases. The benefit of using element-wise feedback is that the RIS phase shifts can be accommodated to fit any type of channel and use case. In scenarios in which there is no correlation between the RIS elements (such as Rayleigh fading), %
an element-wise quantization is needed to guarantee small SNR losses. 
The downside of element-based feedback is that $Nb$ bits are needed to transmit the configuration, which scales linearly with the number of RIS elements.
When there is a correlation between the optimal phase shifts, it can be used to design a more efficient codebook, as shown earlier. However, if element-wise feedback is anyway used, the result in Lemma~\ref{lemma1} explains the array gain loss. 

In Fig.~\ref{fig:Rice_fading}, we show the SNR that is achieved over a Rician fading channel using either the proposed codebook or element-wise quantization. We consider $N=128$ RIS elements and the number of feedback bits is shown on the horizontal axis. In the figure, each point is computed as the average SNR over 10\,000 Monte Carlo trials, with the parameters provided in Table~\ref{tab:table1}. 
The solid curve shows the performance of the proposed codebook. The HPBW condition in \eqref{eq:b_HPBW} is satisfied for $l=9$ feedback bits and is marked by a $\times$-marker. The SNR loss is minor at this point, even if a Rician channel is considered.
On the other hand, the element-wise quantization achieves a significantly lower SNR than the maximum if $b \leq 2$, and almost the maximum SNR if $b=3$. However, for an RIS with $N=128$ elements, that corresponds to $t=384$ bits. Note that these operating points are shown at the right edge of the figure but are actually much further to the right in a figure with linearly-scaled $x$-axis.

\section{Codebook Design with a Static Path}
\label{sec:with-static-path}
In this section, we will also incorporate the static path $\mathbf{h}_s$ into the problem formulation and extend the proposed codebook design. We assume this to be a non-LoS that can be modeled as Rayleigh fading with $\mathbf{h}_s \sim \CN(\mathbf{0},\rho \mathbf{I}_K )$, where $\rho>0$ is the corresponding channel gain. It is crucial to ensure that the RIS phase shifts align with those of $\mathbf{h}_s$ so the paths via the RIS do not cancel the static path. This alignment is achievable by using an additional degree of freedom: a common rotation of all RIS phase shifts to better align with $\mathbf{h}_s$. Specifically, we introduce a common phase shift across all elements of the previous phase-shift matrix as
\begin{equation} \mathbf{D}_{\psi,\phi}=e^{j\phi}\mathbf{D}_\psi. 
\end{equation}
This does not affect the absolute channel gain from the RIS paths, as all elements experience the same phase rotation $e^{j \phi}$. 

Let us once again consider the pure LoS case $(\kappa \to \infty)$ and notice that the optimal phase shift is%
\begin{equation}
    \phi = \arg (\mathbf{a}_K^H(\varphi)\mathbf{h}_s).
    \label{eq:dir_angle}
\end{equation}
With this adjustment, we can  compute the expected value of the maximal channel gain as
\begin{align}
     \mathbb{E}\left\{\Vert\mathbf{h}\Vert^2\right\}&=\mathbb{E}\left\{\left \Vert \mathbf{h}_s + \sqrt{\beta_r \beta_t}\mathbf{a}_K(\varphi)  \mathbf{a}_N^T(\theta_1) \mathbf{D}_{\psi,\phi} \mathbf{a}_N(\theta_2) \right\Vert^2\right\} \nonumber \\
    &=\beta_r\beta_t\left\Vert\mathbf{a}_K(\varphi) \mathbf{a}_N^T(\theta_1) \mathbf{D}_{\psi,\phi} \mathbf{a}_N(\theta_2)\right\Vert^2+ \mathbb{E}\{\|\mathbf{h}_s\|^2\} \nonumber \\
    &\quad+\mathbb{E}\left\{2\Re\left(  \sqrt{\beta_r \beta_t} \mathbf{a}_N^H(\theta_2) \mathbf{D}^H_{\psi,\phi} \mathbf{a}_N^*(\theta_1)  \mathbf{a}_K^H(\varphi)\mathbf{h}_s  \right) \right\}\nonumber \nonumber \\
    &= N^2K\beta_r\beta_t + K\rho \nonumber  \\
    &\quad + 2N\sqrt{\beta_r \beta_t}\mathbb{E}\left\{\Re\left( \mathbf{a}_K^H(\varphi) \mathbf{h}_s e^{-j \phi} \right) \right\} \label{eq:akHhs}    \\
    &= N^2 K \beta_r \beta_t + K \rho + 2N\sqrt{\beta_r \beta_t}\mathbb{E}\left\{\left|\mathbf{a}_K^H(\varphi) \mathbf{h}_s\right|\right\}  \nonumber \\
    &=N^2 K \beta_r \beta_t  + K \rho + N\sqrt{\pi \beta_r \beta_t K \rho} , 
\end{align}
where the last equality follows from the fact that $|\mathbf{a}_K^H(\varphi) \mathbf{h}_s|$ is Rayleigh distributed with $\mathbb{E}\{ |\mathbf{a}_K^H(\varphi) \mathbf{h}_s|\} = \frac{\sqrt{\pi}}{2}\sqrt{K \rho}$. 

Since the control channel only supports a limited number of bits, the phase-shift information $\phi$ regarding the static path must also be quantized to a few bits of resolution. The phase shift is uniformly distributed because $\mathbf{h}_s$ is circularly symmetric. The quantization of $\phi$ leads to a misalignment between the static path and RIS phase shifts, which affects the cross term as explained by the following lemma.


\begin{lemma} \label{th:sinc}
When the phase shift of the static path in \eqref{eq:dir_angle} is selected as $e^{j\phi} = Q_d \left(e^{j\arg (\mathbf{a}_K^H(\varphi)\mathbf{h}_s)}\right)$ with $2^d$ equally-spaced quantization points on the unit circle, where $Q_d(\cdot)$ is the $d$-bit quantization operator that maps the complex number to the nearest quantized value on the unit circle, the average value of the cross term in \eqref{eq:akHhs} becomes 
\begin{align}
  & 2N\sqrt{\beta_r \beta_t}\mathbb{E}\left\{\Re(\mathbf{a}_K^H(\varphi) \mathbf{h}_s e^{-j \phi})\right\} \nonumber\\
   &= \begin{cases} \sinc\left(\frac{1}{2^d}\right) N\sqrt{\pi \beta_r \beta_t K \rho} & \text{ if $d \in \mathbb{Z}_+$}, \\
0 & \text{ if $d=0$}.\end{cases} 
\end{align}

\end{lemma}

\begin{IEEEproof}
 The quantization error $\tilde{\phi}$ for any phase-shift value $\phi$ is uniformly distributed over $\left(-\pi/2^d,\pi/2^d\right)$ for $2^d$ equally-spaced quantization points on the unit circle. We then obtain  
\begin{align}
   &\mathbb{E}\left\{\Re\left(\mathbf{a}_K^H(\varphi) \mathbf{h}_s e^{-j \phi}\right)\right\} = \mathbb{E}\left\{\left|\mathbf{a}_K^H(\varphi) \mathbf{h}_s\right|\cos \tilde{\phi}\right\}\nonumber\\
   &=\sinc\left(\frac{1}{2^d}\right) \mathbb{E}\left\{\left|\mathbf{a}_K^H(\varphi) \mathbf{h}_s\right|\right\}=\sinc\left(\frac{1}{2^d}\right)\frac{\sqrt{\pi}}{2}\sqrt{K \rho},
\end{align}
where the second equality follows from $ \mathbb{E}\{ \cos \tilde{\phi}\}= \mathbb{E}\{ e^{j \tilde{\phi}}\}=\sinc \left(\frac{1}{2^d}\right)$ as in \eqref{eq:MGF}.
\end{IEEEproof}

To maximize the SNR, conveying information about the static path is important, especially if its strength is similar to the RIS-aided path. Lemma 2 shows that the cross term is positive but diminishes as fewer bits are used, thereby reducing the SNR. The term vanishes entirely if $d=0$.

\begin{figure}[t!] \centering
\centerline{\includegraphics[trim=8 2 25 15,clip,width=0.99\columnwidth]{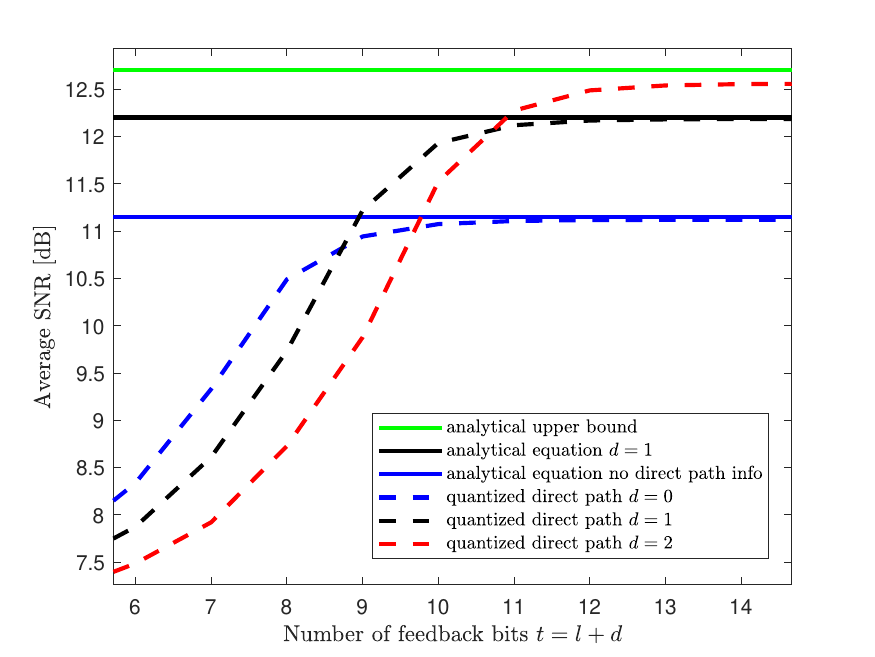}}  \vspace{-4mm}
\caption{The uplink SNR that is achieved with an RIS with 128 elements under pure LoS conditions when using $d$ bits to align the phase shifts of the RIS, with the static paths.}
\vspace{-3mm}
\label{fig:dirpath}
\end{figure}

In Fig.~\ref{fig:dirpath}, we plot the achieved SNR when using a total of $t=l+d$ bits in the control channel for conveying the desired RIS configuration $\mathbf{D}_{\psi,\phi}$: $l$ for $\mathbf{D}_\psi $ and $t$ for $\phi$. We consider the case when there is no static path information in the feedback ($d=0$ bits) and when $d=1$ or $d=2$. The figure shows that providing extra information about the static path increases the SNR. Equation \eqref{eq:akHhs} is also verified in Fig.~\ref{fig:dirpath}. The RIS phase-shift matrix is given by the closest matching entry in \eqref{eq:precoder}. Mathematically, $\mathbb{E}\{\Re\{\mathbf{a}_K^H(\varphi) \mathbf{h}_s e^{-j \phi}\}\} \leq \mathbb{E}\{|\mathbf{a}_K^H(\varphi) \mathbf{h}_s|\}$, with equality only if $\phi$ is not quantized. 
Lemma 2 shows that quantization error reduces this cross term by a factor $\sinc(1/2^d)$, but since the term is small compared to the sum of the other terms, the quantization error can be ignored if $d=2$ or higher. In Fig.~\ref{fig:dirpath}, we notice that when the total number of feedback bits is $t \geq 9$, it is beneficial to allocate $d=1$ or $d=2$ bits to the direct path information, even before fully satisfying \eqref{eq:b_HPBW}, which for the considered $N=128$ elements is guaranteed if $l \geq 9$ or larger. %

\begin{table}[h!]
  \begin{center}
    \caption{Simulation Parameters}
    \label{tab:table1}
    \vspace{-2mm}
    \begin{tabular}{|l|r|} 
     \hline
      \textbf{Parameter} & \textbf{Value} \\
      \hline
      RIS to BS channel gain: $\beta_r$ & $-80$\,dB \\
      UE to RIS channel gain: $\beta_t$ & $-80$\,dB \\
      UE to BS channel gain: $\rho$ & $-120$\,dB  \\
      BS antennas: $K$ & $4$ \\
      RIS elements: $N$ & $128$  \\
      Bandwidth: $B$ & $20$\,MHz  \\
      UE transmit power: $P$ & $100$\,mW  \\
      Receiver noise variance: $\sigma^2$ & $-100.9$\,dBm \\
      RIS to BS angle: $\theta_1$ & $\sim \mathcal{U}(-\pi/2,\pi/2)$ \\
      RIS to UE angle: $\theta_2$ & $\sim \mathcal{U}(-\pi/2,\pi/2)$ \\
      BS array reception angle: $\varphi$ & $\sim \mathcal{U}(-\pi/2,\pi/2)$ \\
       \hline
    \end{tabular}
    \vspace{-6mm}
  \end{center}
\end{table}

\section{Conclusion}

In this paper, we investigated how to feed back the preferred phase-shift configuration from the BS to the RIS in RIS-aided communication systems. The feedback is transmitted over a control channel that supports a limited number of bits.
 We proposed a new codebook-based method that requires a number of bits proportional to the logarithm of the number of RIS elements, while guaranteeing a minimal SNR loss over LoS channels.
 The inclusion of the static path between the BS and the UE is accounted for by adding one or two extra bits, independently of the number of RIS elements. As a rule of thumb, for LoS scenarios (including Rician fading with a practically large $\kappa$ factor), combining \eqref{eq:b_HPBW} and Lemma 2, a total of $t=\log_2(N)+4$ bits ensures that the receiving BS is within the HPBW of the reflected beam from the RIS.%

The proposed codebook-based methods may suffer from performance limitations in handling unstructured randomness in the channel. In contrast, element-wise quantization schemes offer greater adaptability, though with a control overhead that scales linearly with the number of RIS elements, which can become prohibitively large in practical scenarios. Our simulations validated the theoretical findings and illustrated how a reduced size feedback load leads to a quantifiable SNR loss because the array gain diminishes if the phase quantization error exceeds the HPBW.

Our work reinforces the point that an RIS is most suitable in scenarios where the RIS has a LoS path to both the BS and the UE, and can be configured to create a so-called virtual LoS path. Previous works have shown that the relative performance improvements are largest in these scenarios \cite{bjornson2022reconfigurable} and the CSI acquisition becomes much easier since it is sufficient to estimate the LoS angles. In this paper, we show the benefit of this deployment scenario also in terms of requiring a fraction of the feedback signaling over the control channel.

\bibliographystyle{IEEEtran}

\bibliography{IEEEabrv,Referenser}

\end{document}